\documentclass[12pt]{article}
\pdfoutput=1
\usepackage{xcolor}
\usepackage{hyperref}
\usepackage{subcaption}
\usepackage[utf8]{inputenc}

\usepackage{setspace}
\usepackage{amsmath, amssymb, amsthm, float, graphicx}
\numberwithin{equation}{section}

\hypersetup{colorlinks=false, linkcolor=blue, citecolor=red}

\def\beq{\begin{eqnarray}}\def\eeq{\end{eqnarray}}
\def\be{\begin{equation}}\def\ee{\end{equation}}
\def\g{\gamma}
\def\r{\rho}
\def\s{\sigma}
\def\m{\mu}
\def\n{\nu}
\def\a{\alpha}
\def\e{\epsilon}

\def\b{\beta}
\def\d{\delta}

\def\D{\Delta}
\def\G{\Gamma}
\def\l{\lambda}
\def\pd{\partial}

\def\ta{\tau}

\def\la{\langle}
\def\ra{\rangle}

\def\mo{{\mathcal{O}}}

\def\G{\Gamma}

\begin{document}
\title{\bf On critical exponents without Feynman diagrams  }
\date{}

\author{Kallol Sen\footnote{kallol@cts.iisc.ernet.in} ~and Aninda Sinha\footnote{asinha@cts.iisc.ernet.in}\\ ~~~~\\
\it Centre for High Energy Physics,
\it Indian Institute of Science,\\ \it C.V. Raman Avenue, Bangalore 560012, India. \\}
\maketitle
\vskip 2cm
\abstract{In order to achieve a better analytic handle on the modern conformal bootstrap program, we re-examine and extend the pioneering 1974 work of Polyakov's, which was based on consistency between the operator product expansion and unitarity. As in the bootstrap approach, this method does not depend on evaluating Feynman diagrams. We show how this approach can be used to compute the anomalous dimensions of certain operators in the $O(n)$ model at the Wilson-Fisher fixed point in $4-\epsilon$ dimensions up to $O(\epsilon^2)$.}

\vskip 5cm {~~~~~~~~~~~~~~~~~~~~~~~~~~~\it AS dedicates this work to the loving memory of his mother}

\newpage

\tableofcontents

\onehalfspacing

\section{Introduction}

The $\epsilon$-expansion was introduced long time ago by Wilson and Fisher and studied extensively \cite{wilson, vicari, kehrein, kleinert} in order to understand critical phenomena. In $4-\epsilon$ dimensions, the critical theory corresponds to the so-called Wilson-Fisher fixed point. The technique of the $\epsilon$-expansion relies heavily on the machinery of Feynman diagrams. This is both a boon and a bane. For the first few orders in $\epsilon$, one can in principle do the calculation by brute force--for example, the critical exponents or anomalous dimensions for certain operators are known upto $\epsilon^5$ \cite{kleinert}. The series itself is asymptotic and needs to be supplemented by resummation techniques. Going beyond the first few orders leads to computing numerous Feynman diagrams which makes this approach cumbersome and clumsy.

Long time ago, after Wilson's pioneering work, Polyakov \cite{polyakov} derived a set of consistency equations by demanding compatibility between the operator product expansion in conformal field theory and unitarity\footnote{A more recent attempt to study consistency from conformal invariance in the $O(n)$ model is by Petkou\cite{petkou}.}. Specifically, he considered the $O(n)$  model. By looking at the discontinuity in a time-ordered four point function (reviewed below), and using dispersion relations, he was able to compute the position space representation (in a certain limit) of the correlation function. We will frequently refer to this as the unitarity approach. This has to be compatible with what arises on using the OPE which we will refer to as the algebraic approach. He found that there are certain terms which arise in the unitarity approach which are absent in the algebraic one. By setting these to zero gave a set of equations constraining the anomalous dimensions and OPE coefficients. The resulting equations in generality turned out to be too difficult to solve. However, Polyakov showed that in $4-\epsilon$ dimensions, the equations can be solved to leading order in $\epsilon$. In order to achieve this he needed to make two important assumptions: a) The anomalous dimension of $\phi_i$ began at $O(\epsilon^2)$ and b) Consistency could be achieved by just retaining two Lorentz spin zero exchange operators the isospin 0 ($I=0$) scalar and the isospin 2 ($I=2$). With these assumptions the $O(\epsilon)$ anomalous dimensions for the isospin scalar and the isospin 2 operators matched with the Feynman diagram approach of Wilson's. 

Recently, Rychkov and Tan \cite{rychkov} have shown how using conformal invariance alone, one can obtain the leading order anomalous dimensions of a large class of operators. This method just uses three point functions and appears to be quite powerful. However, new ideas are needed to extend this to subleading orders without resorting to the diagrammatic calculations. This was one of the main motivations for us to re-examine Polyakov's original calculations to see if it can be used to obtain the subleading terms in $\epsilon$. As noted in \cite{rychkov}, Polyakov's work \cite{polyakov} has never been used in the literature. Such a method can potentially be useful for a new analytical approach to conformal bootstrap. Indeed following this line of reasoning, we have extended the $\e-$expansion to cubic order in \cite{rg}, thereby yielding new results without Feynman diagrams. 

While the work of Rychkov and Tan eliminates\footnote{Replacing instead by the assumption that certain multiplets recombine following the equations of motion as explained in \cite{rychkov}.} the need for assumption (a) listed above, assumption (b) still needs further investigation. For example can one persist with this assumption upto $O(\e^2)$? In the course of examining this issue, we will show that one can also correctly get the $O(\epsilon^2)$ anomalous dimensions of the above operators. This points to the fact that the unitarity based approach may be more powerful than previously realized and could be another useful way to complement the powerful numerical techniques of the modern conformal bootstrap program  \cite{bootstrap, kaplan, zk, alday,  kss, others}.

The paper is organized as follows. In section \ref{1}, we review some of the results existing in the literature about the anomalous dimensions for the operators $I=0$ and the $I=2$ Lorentz scalars in the $O(n)$ models. The next section \ref{2} deals with the algebraic approach. This idea is based on imposing the operator algebra on the position space Green function derived from the unitarity in momentum space. Demanding consistency, leads to some algebraic equations involving the anomalous dimensions which can then be solved order by order in $\e$. Section \ref{3} deals with the construction of a unitary and crossing symmetric four point amplitude in the momentum space. We have taken only two ($I=0,2$) scalar operators of the $O(n)$ to demonstrate consistency as in \cite{polyakov}. Considering a right branch cut in the direct channel, one can build up this amplitude which is both unitary and crossing symmetric. In the next sections, \ref{4} and \ref{5}, we considered the case of mixed four point functions where one of the operators $\phi_i$ has been replaced with $\phi^2\phi_i$. Section \ref{5} deals with the necessary details for the construction of the unitary and crossing symmetric mixed four point amplitude. As we have pointed out, by demanding consistency one can obtain about the leading $\e$ dependence of the anomalous dimensions of the external operators $\phi_i$ and $\phi^2\phi_i$. In section \ref{6}, we comment on the large spin operators. We show that the anomalous dimensions for these operators (for $n=1$) can be recovered from our bootstrap approach by treating the $\phi^2$ operator as the dominating exchange. 

{\bf Notation:} We will use the same notation as in \cite{polyakov}. $d$ will denote the conformal dimension of the exchanged operator (and not the space time dimension!) while $\Delta$ will denote the conformal dimension of the seed scalar operator in the four point function. We will sometimes use $a=4-\e$.

\section{Anomalous dimensions in the $O(n)$ model}\label{1}

Let us review the $O(\epsilon^2)$ results that arise from the diagrammatic approach \cite{wilson}.
In the $O(n)$ theory we denote the scalar by $\phi_i$ with $i=1,2\cdots n$. We work in $4-\epsilon$ dimensions. The anomalous dimension for $\phi_i$ works out to be
\be
\gamma_{\phi_i}=\frac{n+2}{4(n+8)^2} \epsilon^2 +O(\epsilon^3)\,.
\ee
The anomalous dimension of the Lorentz scalar, isospin zero ($O_0=\phi_i \phi_i$) and isospin 2 ($O_2=\phi_{(i}\phi_{j)}-\frac{2}{n}\phi_k\phi_k \delta_{ij}$) operators are given by
\begin{eqnarray}\label{imp1}
\gamma_{O_0}&=& \frac{n+2}{n+8}\epsilon+\frac{n+2}{2(n+8)^3}(13n+44) \epsilon^2+O(\epsilon^3)\,,\\
\gamma_{O_2}&=& \frac{2}{n+8} \epsilon-\frac{(n+4)(n-22)}{2(n+8)^3}\epsilon^2+O(\epsilon^3)\,.
\end{eqnarray}
The anomalous dimension of certain gradient operators are also known. The one of interest in our work is the Lorentz spin-$\ell$, isospin zero case which includes the stress tensor $O_\ell=\phi_i \partial_{\alpha_1}\cdots \partial_{\alpha_\ell} \phi_i-{\rm trace}$
\be\label{tnsr}
\gamma_{O_\ell}=\frac{n+2}{2(n+8)^2}(1-\frac{6}{\ell(\ell+1)})\epsilon^2+O(\epsilon^3)\,.
\ee
Notice that for $\ell=2$ which corresponds to the stress tensor, the anomalous dimension vanishes to the order shown. We will reproduce this result in the large $\ell$ limit using modern techniques.

Now let us turn to the following question. We know that the diagrammatic approach will need a regularization scheme. As a result, it may be expected that a comparison order by order in $\epsilon$ is not meant to give an agreement unless we know the choice of scheme. However, at criticality, the anomalous dimensions are scheme independent. An explicit check up to $O(\e^2)$ is as follows. It is well known that the beta function is scheme independent up to second order. That is\footnote{In this argument, $b_i$'s are independent of $\epsilon$ in any scheme.}, writing $\beta(g)=b_1 g^2+b_2 g^3+\cdots$, $b_1$ and $b_2$ are scheme independent. The fixed point is obtained by solving $\beta(g)+\epsilon g=0$ order by order in $\epsilon$. Thus the location of the fixed point is scheme independent up to $O(\epsilon^2)$. The general form of the anomalous dimension is $\gamma(g)= a_1 g+a_2 g^2+\cdots$. With a different scheme we have $\gamma(\bar g)=\bar a_1 \bar g+\bar a_2 \bar g +\cdots$. At the fixed point we have to $O(\epsilon^2)$, $a_1 g_*+a_2 g_*^2=\bar a_1 g_*+\bar a_2 g_*^2$. Thus $a_1=\bar a_1, a_2=\bar a_2$ and the anomalous dimensions are scheme independent to $O(\epsilon^2)$. The critical exponents in general are expected to be scheme independent at every order\footnote{We thank Slava Rychkov for emphasising this to us.} in $\epsilon$. The argument is that once we substitute $g=g_*$ and get an expression in terms of $\epsilon$, the only way the scheme dependence would come would be through redefinition of $\epsilon$ which is not allowed.

From Polyakov's work the anomalous dimensions of $O_0$ and $O_2$ follow up to $O(\epsilon)$. We will now review this approach. In the course of doing so, we will correct several typos in the printed version of \cite{polyakov} and also extend it to $O(\e^2)$.

\section{The algebraic approach}\label{2}

We will now use the OPE in a certain limit following \cite{polyakov} to derive the position space form of the four point function. We will focus on only the leading spin-0 exchange. Of course, for the most general analysis one needs to consider the full conformal OPE and impose the algebra at the level of the amplitudes. The OPE for the $O(n)$ model is given by,
\begin{align}
\begin{split}
\phi_i(r)\phi_j(0)&=\frac{\d_{ij}}{r^{2\D_\phi}}+f_0 \frac{\d_{ij}}{r^{2\D_\phi-\D_0}}\mo^0+ f_2 \frac{1}{r^{2\D_\phi-\D_2}}\mo^2_{ij}+\cdots\,,
\end{split}
\end{align}
where the $\cdots$ terms represent the higher order terms either with higher spin or with higher number of fields or both. We will now consider the four point function of the form,
\be
G_{ijkl}(r,r',R)=\la \phi_i(r)\phi_j(0)\phi_k(R)\phi_l(R+r')\ra\,.
\ee
In general $G_{ijkl}(r,r',R)$ will have all the components of the trace, symmetric-traceless and the antisymmetric-traceless parts based on the above decomposition of the OPE. However, for the spin-0 case of interest we have
\be
G_{ijkl}(r,r',R)=\d_{ij}\d_{kl}G_T(r,r',R)+(\d_{ik}\d_{jl}+\d_{il}\d_{jk}-\frac{2}{n} \d_{ij}\d_{kl})G_{ST}(r,r',R)\,,
\ee
where these functions are given by,
\begin{align}
\begin{split}
G_T(r,r',R)&=(rr')^{-2\D_\phi}+f_0{}^2 (rr')^{\D_0-2\D_\phi}R^{-2\D_0}+\cdots\,,\\
G_{ST}(r,r',R)&=f_2{}^2 (rr')^{\D_2-2\D_\phi}R^{-2\D_2}+\cdots\,.
\end{split}
\end{align}
The trace and the symmetric-traceless parts in the Green's function will contain the contributions from both $\mo^{(0)}$ and $\mo^{(2)}$. The main point is to compare the power dependence with what follows from the unitarity approach (next section).

The following point does not appear to be spelt out in \cite{polyakov}. We are interested in the Lorentzian signature\footnote{The analytical continuation from the Lorentzian signature to the Euclidean signature does not destroy the factorization principle. Hence we can continue to the Euclidean signature for doing the integrals in appendix \ref{A0}.} since we are considering time-ordered correlation functions.  We write $x^2=-x^+ x^-+x_ix_i$ and set $x_i=0$ where $x$ is either $r$ or $r'$. Then we consider $x^+\rightarrow0$ and $x^-\rightarrow0$ as otherwise the OPE would be an expansion in terms of twists and operators with the same twists would be expected to be equally important. For further discussions see \cite{zk, alday}. As in \cite{polyakov} we will consider $R^2\gg r^2,r'^2$.

\section{The unitarity based approach}\label{3}

\begin{figure}
\begin{center}
\includegraphics[width=0.8\textwidth]{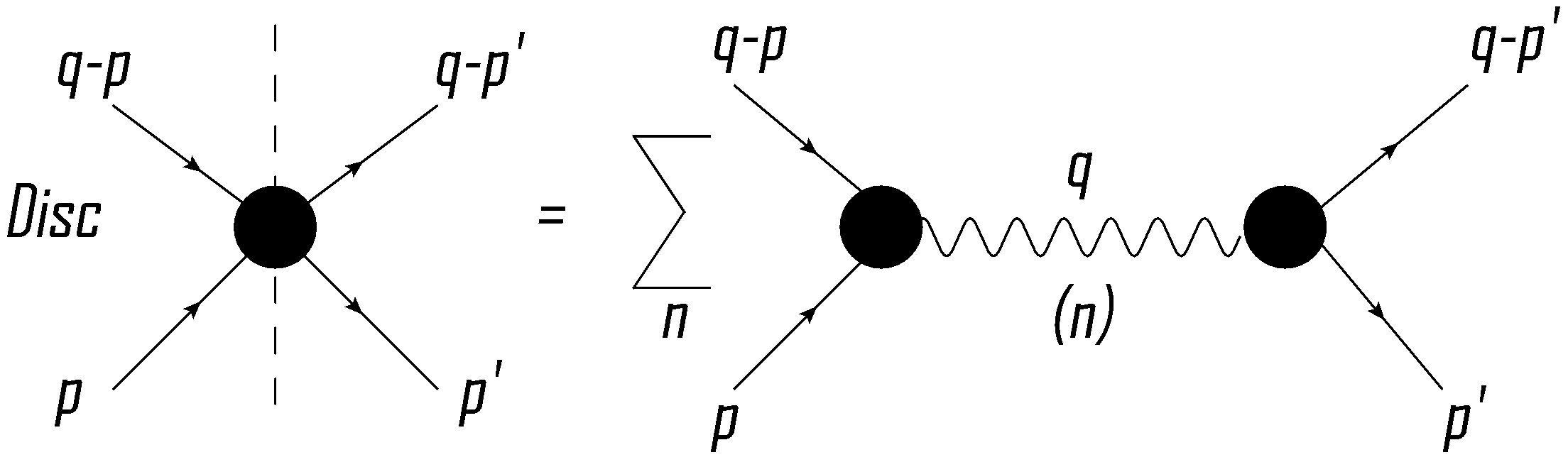}
\caption{A generic time-ordered four point function in momentum space. The left hand side represents the discontinuity in the four point function which can be written as a product of the three point functions with the imaginary part of the exchange propagator. }
\label{fig:scatteringplot}
\end{center}
\end{figure}

We will now briefly summarize the unitarity conditions arising in \cite{polyakov}. The logic is motivated by what arises in scattering theory. We consider time-ordered correlators. In the case of non-conformal QFT, there are various intermediate states which are distinguished by the respective masses and are independent exchanges and have exponentially suppressed (mass dependent) contributions to the correlation function. For the conformal QFT ($m\rightarrow0$), a different kind of intermediate states characterized by the action of operators $\mo_\D$ acting on the vacuum state arise, with a power law contribution to the correlation functions. A discontinuity in an amplitude occurs when intermediate states ($|\mo_\D\ra=\mo_\D|0\ra$) go onshell-when this happens, the amplitude factorizes. In momentum space, the discontinuity arises from the imaginary part of the Feynman propagator as shown in figure \ref{fig:scatteringplot}. Thus the discontinuity (in the complex momentum of the exchanged operator) in a four point function can be written as the product of three point functions times the imaginary part of the propagator. By knowing the discontinuity in the complex momentum space, we can use dispersion relation techniques to know the four point function everywhere. Then we can Fourier transform back to position space and compare with what arises in the algebraic approach. 

In a conformal field theory, the three point function is fixed up to some overall constants. We start by considering the four point function of scalars. As in \cite{polyakov}, we begin by considering only Lorentz scalar exchange. 
 Thus if we define the three point functions as $T_d(q,p,p')$ where $d$ is the dimension of the exchanged operator and $\text{Im} D(q^2)$ as the imaginary part of the two point function, then, the discontinuity of the fourier transform of \footnote{The discontinuous part follows the factorization in the momentum space.}the four point amplitude is given by,
\be
\text{F.T.~Disc}\la T\phi(r)\phi(0)\phi(R)\phi(R+r')\ra=T_d(p,q)\ \text{Im} D(q^2)\ T_d(q,p')\,.
\ee
Due to scale invariance, the imaginary part of the propagator takes the form,
\be\label{imd}
\text{Im}D(q^2)=\text{const.}\ (q^2)^{d-a/2}\,,
\ee
where
$$
a=4-\epsilon\,.
$$
Now we consider the $O(n)$ model with the field $\phi_i$. As in \cite{polyakov}, we assume that consistency between the unitarity approach and the algebraic approach can be obtained using the operators (bilinear in the fields) with Lorentz spin zero and isospins 0 and 2. The s-channel discontinuity is written as
\begin{eqnarray}
\text{Disc}_q \frac{A_{iklm}(q,p,p')}{p^2 (p-q)^2 p'{}^2(q-p')^2}=&& a_0 \delta_{ik}\delta_{lm} T^{(0)}(q,p)T^{(0)}(q,p')\text{Im}D^{(0)}(q)\nonumber\\&&+a_2 B_{iklm}T^{(2)}(q,p)T^{(2)}(q,p')\text{Im}D^{(2)}(q)\,,
\end{eqnarray}
where $B_{iklm}=(\delta_{il}\delta_{km}+\delta_{lm}\delta_{kl}-\frac{2}{n}\delta_{ik}\delta_{lm})$ and where the antisymmetric rank-4 $O(n)$ tensor does not play a role since it involves the exchange of odd Lorentz spin. Here the superscripts $(0), (2)$ represent the isospin of the exchanged operator.  $a_0$ and $a_2$ will turn out to be related to the OPE coefficients. The definition of $A_{iklm}$ concurs with the definition of the discontinuous part of the amplitude in eq. 5.1 of \cite{polyakov}.

\vspace{0.5cm}
\noindent\textbf{ Limiting values of $T_d(p,q)$:}
\vspace{0.2cm}

Unlike appendix A of \cite{polyakov} where the expressions are worked out for $4$ dimensions, we will always be in $4-\e$ dimensions. We denote $q^2=s, p^2=v_1, (p-q)^2=v_2, (p')^2=w_1, (p'-q)^2=w_2$ and following \cite{polyakov} consider $v_1\approx v_2\equiv v$ and $w_1\approx w_2=w$.
The limiting values of the above three point function in the limits when,
\be
s(q^2)\ll v(p^2)\,, \ \ \text{and} \ \ s(q^2)\gg v(p^2)\,.
\ee
We are assuming that $(p-q)^2\approx p^2$. Thus in the two limits the three point function takes the asymptotic form,
\be\label{tdl}
   T_d(p,q) =\frac{1}{v^2} \left\{
     \begin{array}{lr}
      f_1 v^{-d/2+\D+\e/2} & : s\ll v\\
      f_2 s^{2-d/2-\D-\e/2}v^{-2+2\D+\e} & : s\gg v
     \end{array}
   \right.
\ee 
where $\text{Im}D(s)=s^{d-a/2}$ where $a=4-\e$. Further the coefficients $f_1$ and $f_2$ take the form,
\be
f_1=L_{d,d,2\D-d}\frac{\G(d/2)^2}{\G(d)}\,,  \ f_2=L_{d,d,2\D-d}\frac{\G(2-\D-\e/2)^2\G(-1+d+\e/2)}{\G(1+d/2-\D)\G(2+d/2-\D-\e/2)}\text{Re}[(-1)^{\frac{1}{2}(-d-2\D-\e)}]\,,
\ee
where $L_{d,d,2\D-d}$ can be read off from \eqref{NL}. We will redefine the $a_I$ to absorb these $L_{d,d,2\D-d}$ in them and ignore these overall coefficients in the subsequent discussions.  Similar expressions are found involving $w$ instead of $v$. Knowing the discontinuity, we can now compute the amplitude using the dispersion relation
$$
A(s,v,w)=\int_0^\infty \frac{ds'}{s'-s} \text{Disc}_{s'} A(s',v,w)\,.
$$
 Furthermore, we will make the amplitude manifestly crossing symmetric by adding the $t$ and $u$ channel contributions.

Note that this is not the usual use of the dispersion relation where we add the s and u channels making use of the constraint $s+t+u=0$. This procedure is not manifestly crossing symmetric. In fact using a crossing symmetric form may lead to the worry that we are over-counting the contribution of the physical states. However, we will explicitly show that in our approximations, the physical state contribution comes only from the s-channel and hence there is no overcounting.

\subsection{Calculation of the four point functions}

Using \eqref{imd} and the limiting values of $T_d(p,q)$ given in \eqref{tdl} we can now calculate the four point function in a regime where $v(p^2)\gg w(p'^2)\gg s(q^2)$. We will discuss the regimes separately in $s$, $t$ and the $u$ channel. Note that whereas in the $s$ channel the exchange particle has momentum $q$, the corresponding momentum for the exchanges in the $t$ and the $u$ channel is $p-p'$ and $p+p'-q$ respectively. The full interacting and crossing symmetric amplitude becomes a sum of the individual contributions of the amplitude from the three channels. We have
\be
A(q,p,p')=A^{(s)}(q,p,p')+A^{(t)}(q,p,p')+A^{(u)}(q,p,p')\,.
\ee

\subsubsection{$s$-channel}   

The amplitude in the $s$-channel is given by,
\begin{align}
\begin{split}
A(s,v,w)\approx&\frac{f_1{}^2}{(v w)^{d/2-\D-\e/2}}\int_s^w \frac{ds'}{s'}s'^{d-a/2}+\frac{f_1f_2 w^{-2+2\D+\e}}{v^{d/2-\D-\e/2}}\int_w^v \frac{ds'}{s'}s'^{2+d/2-a/2-\D-\e/2}\nonumber\\
&+f_2{}^2(v w)^{-2+2\D+\e}\int_v^\infty \frac{ds'}{s'}s'^{4-a/2-2\D-\e}\,.
\end{split}
\end{align}
We have taken the leading term in the two limits $s\ll v$ and $s\gg v$ for the discontinuous part of the three point functions $T_d(p,q)$. One might ask about the relevance of the subleading terms we have neglected in the analysis. In the analysis and the regime we are interested in, these subleading terms do not interfere\footnote{For $0<s'\ll s$ the denominator cannot be written as $\frac{1}{s'-s}\approx \frac{1}{s'}$, so this regime will not interfere with the terms here.}. Till now we have not put any index on the conformal dimensions $d$ implying that this form is general for the exchanges we have considered. 
The final expression is,
\begin{align}
\begin{split}
A^{(s)}(q,p,p')&=g_1q^{2d-a}(pp')^{2\D+\e-d}+g_2 p^{2\D+\e-d}p'^{2\D+\e+d-a}+g_3 p^{4-a}p'^{-4+4\D+2\e},
\end{split}
\end{align}
where the functions, $g_i$ take the form,
\begin{align}\label{gi}
\begin{split}
g_1&=-\frac{f_1{}^2}{d-a/2}\,, \ \ g_2=\frac{f_1{}^2}{d-a/2}-\frac{f_1f_2}{2+d/2-a/2-\D-\e/2}\,,\\
g_3&=\frac{f_1f_2}{2+d/2-a/2-\D-\e/2}-\frac{f_2{}^2}{4-a/2-2\D-\e}\,, \ \ g_4=\frac{f_2{}^2}{4-a/2-2\D-\e}\,.
\end{split}
\end{align}
The $f_i$ and $g_i$ written above will come with an additional label $I$ depending on whether the exchange in $\phi_i \phi_i$ for $I=0$ or $\phi_i\phi_j+\phi_j\phi_i-\frac{2}{n} \phi_i\phi_i$ for $I=2$. 

\subsubsection{$t$-channel}

In the $t$ channel the exchange momentum is $s=(p-p')^2$. The ingoing and the outgoing momenta are $p,q-p$ and $p',q-p'$ respectively. Thus the arguments of $T_d(q,p)\rightarrow T_d(p-p',p-q)$. All the other calculations remain the same. One other important difference is that since here we are in the regime $p^2\gg p'{}^2\gg q^2$, hence the exchange momentum is always atleast $O(v)$. Thus,
\be
A(s,v,w)=\int \frac{ds'}{s-s'}A(s',v,w)\approx \int_v^\infty \frac{ds'}{s'}A(s',v,w)\,,
\ee
and the lower limits do not apply here. Again using the form of the propagator in \eqref{imd} and the limiting form of $T_d(q,p)$ in this channel we get,
\be
A^{(t)}(q,p,p')=-\frac{1}{p^4p'^4}g_4 p^{4-a}p'^{-4+4\D+2\e}\,.
\ee
\subsubsection{$u$-channel}
Similar to the $t$-channel we have an exchange momentum $s=(p+p'-q)^2$ for which we have an analogous expression. This is again,
\be
A^{(u)}(q,p,p')=-\frac{1}{p^4p'^4}g_4 p^{4-a}p'^{-4+4\D+2\e}\,.
\ee

\subsubsection{The full interacting amplitude} 
Upto now we have not used the precise form of the exchanged operator. As in \cite{polyakov}, we will now assume that the exchanged operators are just $O^{(0)}=\phi_i\phi_i$ and $O^{(2)}=\phi_i\phi_j+\phi_j\phi_i-2/n O^{(0)}\d_{ij}$ scalars.  Thus depending on what among these operators is getting exchanged the amplitudes all the channels will come with an index $I$ and moreover the contributions of the $t$ and the $u$ channels will come dressed with some mixing between these exchanges. We derive these mixing coefficients in appendix \eqref{mixapp}. With these mixing terms, we can write down the full interacting amplitude (with the index $I$ denoting the particular exchange) for the $\la\phi_i\phi_j\phi_k\phi_l\ra$ as,
\be
A_{ijkl}(q,p,p')=\d_{ij}\d_{kl}A^0(q,p,p')+(\d_{ik}\d_{jl}+\d_{il}\d_{jk}-\frac{2}{n}\d_{ij}\d_{kl})A^2(q,p,p')\,,
\ee
where,
\be
A^I(q,p,p')=a^I[g_1^I q^{2d^I-a}(pp')^{2\D+\e-d^I}+g_2^I p^{2\D+\e-d^I}p'^{2\D+\e+d^I-a}]+(a^Ig_3^I-b_I)p^{4-a}p'^{-4+4\D+2\e}\,,
\ee
and further,
\be\label{bi}
b_I=\sum_{J=0,2}c_{IJ}a_J g_4^J\,.
\ee
$d^I$ is the conformal dimension of the exchanged scalar for $I=0,2$ respectively and $c_{IJ}'$s are given in \eqref{mixcoeff}.  

\subsection{Coordinate space Green function}

The next step is to use the four point Green function and Fourier transform it to the coordinate space so that we can use the operator algebra. We begin by writing,

\be
\la T\phi_i(r)\phi_i(0)\phi_k(R)\phi_l(R+r')\ra=G(r,r',R)=\d_{ij}\d_{kl}G^0(r,r',R)+(\d_{ik}\d_{jl}+\d_{il}\d_{jk}-\frac{2}{n}\d_{ij}\d_{kl})G^2(r,r',R)\,.
\ee

On the other hand we can also write,
\begin{align}
\begin{split}
\la T\phi_i(r)\phi_i(0)\phi_k(R)\phi_l(R+r')\ra&=\d_{ij}\d_{kl}(rr')^{-2\D_\phi}+(\d_{ik}\d_{jl}+\d_{il}\d_{jk})R^{-4\D_\phi}\\
&+\int \frac{d^{4-\e}pd^{4-\e}p'd^{4-\e}q}{p^2p'^2(p-q)^2(p'-q)^2}e^{ipr+ip'r'+iqR}A_{ijkl}(q,p,p')\,,
\end{split}
\end{align}
where $A_{ijkl}(q,p,p')$ also has a similar decomposition in terms of the tensor structure of $O^{(0)}$ and $O^{(2)}$ scalar exchanges. Comparing the above two expressions we get,
\begin{align}
\begin{split}
G^0(r,r',R)&=(rr')^{-2\D_\phi}+\frac{2}{n}R^{-4\D_\phi}+\int \frac{d^{4-\e}pd^{4-\e}p'd^{4-\e}q}{p^2p'^2(p-q)^2(p'-q)^2}e^{ipr+ip'r'+iqR}A^0(q,p,p')\,,\\
G^2(r,r',R)&=R^{-4\D_\phi}+\int \frac{d^{4-\e}pd^{4-\e}p'd^{4-\e}q}{p^2p'^2(p-q)^2(p'-q)^2}e^{ipr+ip'r'+iqR}A^2(q,p,p')\,.
\end{split}
\end{align}

In the regime of interest $p^2\gg p'^2\gg q^2$ or equivalently in the coordinate space $r^2, r'^2\ll R^2$, we can to the leading order drop the oscillating exponentials and adjust the limits of the integrals\footnote{We are taking the integral measure as $d^{4-\e}x\propto x^{3-\e}dx$ for $x\in \{p,p',q\}$.} over $p$ and $p'$ to
$q\ll p'\ll p\ll r^{-1},r'^{-1}$. Further we can replace the limits of the $q$ integral from $0$ to $R^{-1}$. Thus the Fourier transform takes the form,
\begin{align}
\begin{split}
\int \frac{d^{4-\e}pd^{4-\e}p'd^{4-\e}q}{p^2p'^2(p-q)^2(p'-q)^2}e^{ipr+ip'r'+iqR}A^I(q,p,p')& \approx \int_0^{R^{-1}} d^{4-\e}q\times\\
&\int_{R^{-1}}^{r'^{-1}} \frac{dp'}{p'}p'^{-\e}\bigg(\int_{p'}^{1/r}+\int_{p'}^{1/r'}\bigg)\frac{dp}{p}p^{-\e}A^I(q,p,p')\,.
\end{split}
\end{align}

Using the expressions for the integrals given in the appendix \eqref{A}, we get,

\begin{align}
\begin{split}
G^0(r,r',R)=&(rr')^{-2\D_\phi}+\frac{2}{n}R^{-4\D_\phi}+a_0[g_1^0 S_{2d^0-a,2\D+\e-d^0,2\D+\e-d^0}(r,r',R)\\
&+g_2^0 S_{0,2\D+\e-d^0,2\D+\e+d^0-a}(r,r',R)]+(a_0g_3^0-b_0)T_{0,-4+4\D+2\e}(r,r',R)\,,
\end{split}
\end{align}
\begin{align}
\begin{split}
G^2(r,r',R)=&R^{-4\D_\phi}+a_2[g_1^2 S_{2d^2-a,2\D+\e-d^2,2\D+\e-d^2}(r,r',R)\\
&+g_2^2 S_{0,2\D+\e-d^2,2\D+\e+d^2-a}(r,r',R)]+(a_2g_3^2-b_2)T_{0,-4+4\D+2\e}(r,r',R)\,.
\end{split}
\end{align}

We will now assume that the conformal dimensions $\D$ and $d$ for the external operators $\phi$ and the exchange can be given in an $\epsilon$-expansion of the form,
\be
\D=1-\frac{\e}{2}+\g_\phi \e^2\,, \ \ d^{0,2}=2-\e+\d^{0,2}(\e)\,.
\ee 
We will use $\g_\phi=(n+2)/(4(n+8)^2)$ as follows from conformal symmetries of three point functions in \cite{rychkov}. The coefficients of the $\log\bigg(\frac{R^2}{rr'}\bigg)$ terms come only from the $T$ terms in the above expression. Hence setting,
\be
a_I g_3^I-b_I=0\,, \  \  I=0,2\,,
\ee
where $b_I$ is defined in \eqref{bi}, $g_i^I$ defined in \eqref{gi}, gives us two relations for the anomalous dimensions for $\mo_0$ and $\mo_2$. The remaining relations are obtained by matching the coefficients of $R^{-4\D_\phi}$ from the remaining of $G^0$ and $G^2$. The coefficient of $R^{-4\D_\phi}$ takes the form,
\be\label{ope}
a_I\bigg[\frac{g_1^I}{(4+2\d^I)(\d^I+\e-2a\e^2)^2}+\frac{2g_2^I}{(4-\e)(\e-4a\e^2)(\d^I+\e-2a\e^2)}\bigg]= \left\{
     \begin{array}{lr}
     \frac{2}{n} & : I=0\\
      1 & : I=2
     \end{array}
   \right.
\ee
$g^I_i$ etc. can be found in \eqref{gi}. Also $\d^I$ and $a_I$ are assumed to have an expansion of the form,
\be\label{expansion}
\d^I(\e)=\a_I\e+\b_I \e^2+O(\e)^3\,, \ \ a_I(\e)=\r_I\e^3+\s_I\e^4+O(\e^5)\,.
\ee
The reason behind the particular dependence of $a_I\sim O(\e^2)$ is because of the fact that we expect the OPE coefficients to start at $O(1)$.  The solution to the above set of equations is given by,
\be
\a_0=-\frac{6}{n+8}\,, \ \ \a_2=-\frac{n+6}{n+8}\,, \ \ \b_0=\frac{(n+2)(13n+44)}{2(n+8)^3}\,, \ \ \b_2=-\frac{(n+4)(n-22)}{2(n+8)^3}\,.
\ee
Thus the expressions for the dimensions for the scalars $\mo_0$ and $\mo_2$ become,
\begin{align}
\begin{split}
d^0&=2-\e+\frac{n+2}{n+8}\e+\frac{(n+2)(13n+44)}{2(n+8)^3}\e^2+O(\e^3)\,,\\
d^2&=2-\e+\frac{2}{n+8}\e-\frac{(n+4)(n-22)}{2(n+8)^3}\e^2+O(\e^3)\,.
\end{split}
\end{align}
These  match exactly with eq(\ref{imp1})!

We also provide the OPE coefficients to next to the leading order (in $\e$) results pointed out in \cite{polyakov}. The relevant equation for the OPE coefficients is \eqref{ope} where $a_I$ encodes the information of the squares of the OPE coefficients for the corresponding scalar exchange $\mo_I$. Once we have solved for the anomalous dimensions of the scalar exchanges $\d_I(\e)$ from \eqref{expansion}, we can put that information in \eqref{ope} to solve for the leading and next to leading order in $\e$. To extract the leading order, note that the {\it rhs} of \eqref{ope}. Using the normalizations used, we can see that the correct OPE coefficients squared are of the form,
\be
f_I^2=\frac{a_I g_1^I}{(2\D-d_I)^2 2d_I}\,.
\ee
With the above normalizations, the OPE squared coefficients for the exchange $\mo_I$ start with $f_I^2=1+O(\e)+\cdots$. For the exchanges $\mo_0$ and $\mo_2$, the OPE squared coefficients are given by,
\begin{align}
\begin{split}
f_0^2&=\frac{2(n+8)}{n(4-n)}+\frac{40384+2n(11272-n(n+8)(n-150))}{27(n-4)^2n(n+2)(n+8)}\e+O(\e)^2\,,\\
f_2^2&=\frac{8+n}{4+n}+\frac{1}{54}\bigg(66+n+\frac{610}{4+n}-\frac{486}{8+n}\bigg)\e+O(\e)^2\,.
\end{split}
\end{align}

{\it Why did the calculation work to second order?} Here is a plausible argument. The OPE coefficient appears in the calculation as squared. It is reasonable to assume that there are no fractional powers in $\epsilon$ so
that the contributions from other operators that we have not considered will begin at least $O(\epsilon^2)$ higher than in our equations, namely because to $O(\epsilon)$ this approach gave the expected answer which is fixed by other considerations \cite{rychkov}. Thus the contributions from other operators can be expected to start at $O(\epsilon^3)$. For example, in the free theory $\la \phi\phi\phi^2\ra$ is non zero but the higher three point functions $\la \phi\phi\phi^n\ra$ for $n>2$ are zero. To see that this is consistent, it may be necessary to consider mixed correlators as well. For a more rigorous understanding of this, see \cite{rg}. We will begin the analysis of mixed correlators in the next section, leaving a complete analysis for future work.

\section{Mixed correlators}\label{4}

We will consider one more example to illustrate this method. We will consider the example for a mixed correlator: $\la \phi_i(r)\phi_j(0)\phi_k(R)\phi^2\phi_l (R+r')\ra$. The last operator is a composite one made out of three fundamental fields $\phi$. We will treat this in the same way as for the four point function of four fundamental fields $\phi_i$. The only difference in this case is going to be that we will need here two different sets of OPEs in this case \textit{viz.} $\phi_i\times \phi_j$ and $\phi_i\times \phi^2\phi_j$ respectively where in the previous examples we only needed one type of the OPE $\phi_i\times \phi_j$.

\subsection{ OPE for mixed correlators}

 We will consider separately these two OPEs now. For the $\phi_i\times \phi_j$ case we can write down with two scalars $\mo^0$ and $\mo^2$ as follows,
\be
\phi_i(r)\times \phi_j(0)= r^{-2\D_\phi}\d_{ij}{\mathbb{I}}+r^{\D_0-2\D_\phi}f_0\mo^{(0)}\d_{ij}+r^{\D_2-2\D_\phi}f_2\mo^{(2)}_{ij}+\cdots\,,
\ee
whereas for the other OPE we firstly will not have the identity term since the two point function vanishes (they are different operators) and secondly there are additional scalars even in the free field limit that gives a non zero three point function and hence should be considered in the OPE itself. Thus,
\begin{align}
\begin{split}
\phi_i(R)\times\phi^2\phi_j(R+r')&=r'^{\D_0-\D_\phi-\D_3}g_0\mo^{(0)}\d_{ij}
+r'^{\D_2-\D_\phi-\D_3}g_2\mo^{(2)}_{ij}+r'^{\D_{40}-\D_\phi-\D_3}h_0\mo^{(4,0)}_{ij}\\
&+r'^{\D_{42}-\D_\phi-\D_3}h_2\mo^{(4,2)}_{ij}\,.
\end{split}
\end{align}
The scalars $\mo^{(4,0)}$ and $\mo^{(4,2)}$ are the fourth order scalars of $I=0,2$. Note that in the second OPE we have denoted the conformal dimension for the composite operator by $\D_3$. We can as well consider the order four scalars in the OPE of $\phi_i\times \phi_j$ as well. Thus for the four point function we get,
\be
\la \phi_i(r)\phi_j(0)\phi_k(R)\phi^2\phi_l(R+r')\ra=f_0 g_0r^{\D_0-2\D_\phi}r'^{\D_0-\D_\phi-\D_3}R^{-2\D_0}+f_2 g_2r^{\D_2-2\D_\phi}r'^{\D_2-\D_\phi-\D_3}R^{-2\D_2}+\cdots\,,
\ee
where we have neglected the higher order scalars for the time. Let us remind the readers about the notations again. $\D_\phi=1-v\e/2+O(\e^2)$ and $\D_3=3(1-\e/2)+b\e+O(\e^2)$ are the conformal dimensions for $\phi_i$ and composite operator $\phi^2\phi_i$ and $\D_i=2+\d_i$ are the conformal dimensions for the scalars $\mo^{(0)}$ and $\mo^{(2)}$ respectively. Consistency should lead to $b=v=1$ as in \cite{rychkov}.

\subsection{Leading anomalous dimensions for $\phi_i$ and $\phi^2\phi_i$}

Since the contruction of the four point function for the mixed corelator is along the same lines as for the four point function of the fundamental scalars $\phi_i$, we defer the relevant details to appendix \eqref{C} and quote here the final results. Note that \eqref{constraint} serves as the constraint which can be solved order by order in $\e$ to get the anomalous dimensions of the extenal operators. We will assume a form for the dimensions for $\phi_i$ and $\phi^2\phi_i$ as,
\be
\D_\phi=1-v\frac{\e}{2}\,,\ \  \D_{\phi^3}=3\bigg(1-\frac{\e}{2}\bigg)+b\ \e\,.
\ee
The leading order expansion in $\e$ for \eqref{constraint} is supposed to impose constraints on the coefficients $b$ and $v$. The expected answer in $b=v=1$. Moreover we assume for the anomalous dimensions of the $\phi^2$ exchange as,
\be
d_I=2+\r_I\e+\s_I\e^2\,, \ \ \text{and,} \ \  a_I=L_1^IL_2^I(\n_I\e^3+\m_I\e^4+\eta_I\e^5)\,,
\ee
where $L_1^I$ and $L_2^I$ are the overall factors associated with the three point functions $\la\phi\phi\mo\ra$ and $\la\phi(\phi^2\phi)\mo\ra$ given in \eqref{NL}. Using these to expand \eqref{constraint}, we find that the leading order term does not show any divergence as $(b,v)\rightarrow1$ but the subleading terms have double poles as $(b,v)\rightarrow1$ in the form of $((v-1)(1+v-2b))^{-1}$. To remove this divergence we assume that the leading term in $a_I$ goes like,
\be
\n_I=(v-1)(1+v-2b)\xi_I\,, \ \ \text{and} \ \ \m_I=0\,,
\ee
where $\xi_I$ is the part of $a_I$ without the zeros. With this substitution, it is easy to see that not only the divergence is removed at the subleading order but now the leading term in $a_I$ goes to zero as well as we take either $v\rightarrow1$ or $b\rightarrow1$. We have to justify how to see that the leading behaviour of the OPE coefficients are double zeros in $b$ and $v$. Note that the overall factor in $\n_I$ can be expanded to,
\be
(v-1)(1+v-2b)=(b-v)^2-(b-1)^2\,.
\ee
Hence the OPE coefficient for the mixed correlator takes the form,
\be
\text{OPE}_{\phi\phi}\times\text{OPE}_{\phi (\phi^2\phi)}=\frac{((b-v)^2-(b-1)^2)\xi_Ic_1^IL_1^IL_2^I}{2d_I(2\D-d_I)(\D+\D_3-d_I)}=\frac{1}{\e^2}((b-v)^2-(b-1)^2)\xi_I+\cdots\,.
\ee

So if we set $v=1$ first, then this factor vanishes for all $b$ and $b=1$ does not have to be imposed as a required condition. It will imply that the consistency condition does not apriori depend on the leading $\e$ dependence of the dimension for the operator $\phi^2\phi_i$. Instead if we ask the following question,\textit{`` Given that the anomalous dimension for the $\phi^2\phi_i$ operator starts with $O(\e)$ with unit coefficient, what can we say about the anomalous dimension for the operator $\phi_i$?"} So the idea is to first set $b=1$ in the calculation. One can readily see that the overall factor develops a double zero at $v=1$. Thus the OPE expansion begins like,
\be\label{ope1ope2}
\text{OPE}_{\phi\phi}\times\text{OPE}_{\phi (\phi^2\phi)}=\frac{(1-v)^2}{\e^2}\xi_I+\cdots\,,
\ee
where $\cdots$ represent terms with higher orders of $\e$. Since the exchanges $I=0,2$ scalars are present even in the free limits, then one would be forced to conclude that the leading order behaviour of $a_I$ should be $O(1)$ term and higher powers of $\e$. We can see from the above expansion, that the $O(1/\e^2)$ term vanishes only for $v=1$ which is another way of putting that the dimension of the fundamental scalars do not receive corrections at order $O(\e)$. Thus a consistency statement can be formulated as,

\textit{``Given that the anomalous dimensions for the $\phi^2\phi_i$ operator contributes at $O(\e)$, then we will require the anomalous dimensions of the external scalars $\phi_i$ to start at $O(\e^2)$ for a consistent free limit to exist with $i=0,2$ scalar exchanges."}

For completeness, we also compute the coefficient $\eta_i$ in the limit when $(b,v)\rightarrow1$. As expected these quantites are free from poles in the above limit and given by,
\be
\eta_i=\frac{1}{2}\r_I(1+\r_I)(1+2\r_I)^3\sum_{J\neq I}c_{IJ}\xi_J\frac{(1+\r_J)}{(1+2\r_J)^2}\,,
\ee
where $\r_I$ are the $O(\e)$ terms in the anomalous dimensions for the $I=0,2$ scalar exchanges, $c_{IJ}$ are the mixing coefficients and $\xi_I$ are the leading order OPE coefficients. 

Alternatively we could ask the question, {\it ``what happens if we directly started out assuming that the anomalous dimension of the fundamental operator $\phi_i$ is $O(\e^2)$?} There is no apriori reason to assume that the two expansions are related. So to be completely generic we will assume that,
\be
 a'_I=L_1^IL_2^I(\n'_I\e^3+\m'_I\e^4+\eta'_I\e^5)\,.
\ee
Expanding in $\e$ we find that as before $\m'_I=0$ and also that {\it the double zero reflects in the OPE coefficients as a factor of $(b-1)^2$.} Then the analog of \eqref{ope1ope2} is given by,
\be
\text{OPE}_{\phi\phi}\times\text{OPE}_{\phi (\phi^2\phi)}=\frac{(1-b)^2}{\e^2}\xi'_I+\cdots\,,
\ee
where $\n'_I=(1-b)^2\xi'_I$. We can see that for the above product to begin at $O(1)$ we have to put $b=1+\g_3\e^2$ and then the subleading coefficients $\eta'_I$ are related to the previous $\eta_I$ (when we set $b=1$ first) as,
\be
\eta'_0=-\frac{1}{2}\eta_0\,,\ \  \eta'_2=-\frac{1}{4}\eta_2\,.
\ee
It is not difficult to relate the coefficients $\xi_I$ and $\xi'_I$ by comparing with each other upto some overall constants. But the ratio will depend on the the anomalous dimensions of $\phi_i$ and $\phi^2\phi_i$. We did not consider the subleading terms in $\e$ for the anomalous dimensions of $\phi^2\phi_i$ since we expect higher order/spin corrections to play a role here. The reason is that in this case, the OPE coefficients do not appear as squared.

\section{ $\e$-expansion for large spin operators from bootstrap}\label{6}

For completeness, we discuss the anomalous dimensions of large spin operators using modern methods. For $(n=1)$ and

\be\label{hs}
J_{\ell}=\phi \pd_{\m_1}\dots \pd_{\m_\ell}\phi \,,
\ee
the anomalous dimension is given by,
\be\label{anohs}
\g_{J_{\ell}}=\bigg(1-\frac{6}{\ell(\ell+1)}\bigg)\frac{\e^2}{54}+O(\e^3)\,.
\ee
The spin independent part comes from the anomalous dimensions of the scalar $\phi$. The spin dependent part is given by,
\be
\g_{J_\ell}=-\frac{\e^2}{9\ell^2}\,.
\ee
We will now reproduce this from the results of \cite{kaplan}. Since we know that the approximate conformal blocks do not depend on the dimension $d$ \cite{kaplan, kss}, we can use them to solve for the leading anomalous dimensions for large $\ell$ operators with twists $\ta=2\D_\phi$ \cite {kaplan, kss}--we use the normalizations in \cite{kaplan},
\be\label{g0}
\g(0,\ell)=\frac{\g_0}{\ell^{\ta_m}}\,, \ \text{where}\ \g_0=-\frac{2P_m\G(\D_\phi)^2\G(\ta_m+2\ell_m)}{\G(\D_\phi-\frac{\ta_m}{2})^2\G(\frac{\ta_m}{2}+\ell_m)^2}\,.
\ee
Considering the case where the scalar $\phi^2$ is the dominant exchange, we can set,
\be
\ta_m=d-2+\e \g_{\phi^2}\,.
\ee
The scalar $\phi$ also acquires anomalous dimension,
\be
\D_\phi=\frac{d-2}{2}+\e^2\g_{\phi}\,,
\ee
although these are at a higher order in $\e$ that those of $\phi^2$ operator. Plugging this in \eqref{g0}, setting $\ell_m=0$ and using the mean field $P_m=2$ \cite{kaplan}, we find that,
\be
\g_0=-\frac{\e^2}{9}\,, \ \text{so} \ \g(0,\ell)=-\frac{\e^2}{9\ell^2}+O(\e^3)\,,
\ee
Instead if we put in the form of the $\g_\phi$ for the $O(n)$ models as given in \cite{rychkov}, we should get \eqref{tnsr}. For the stress tensor exchange (or any corresponding higher spin exchange with the same twist), $\ta_m=d-2$. The leading contribution of $\G(\D_\phi-\frac{\ta_m}{2})$ in the denominator in \eqref{g0}, starts with $1/\e^4$ while all the other terms are $O(1)$. Thus the contribution to $\g_0$ from these operators begins at a higher order, namely $\sim \e^4$. 

Clearly this is more subleading than the $\phi^2$ exchange. Also since the higher three point functions of the form $\la \phi\phi \phi^n\ra$ for $n>2$ are all zero in the free limit we expect their contributions to be suppressed further by $O(\e^2)$ compared to the leading $\phi^2$ exchange. With this, one can see that the higher spin operators of the form in \eqref{hs} are indeed given as in \eqref{anohs} to the leading order. It may be possible to use the results of \cite{kss} to get the anomalous dimensions of the large spin and large twist operators. It will also be interesting to see if we can reproduce these results using Polyakov's approach.

\section{Concluding comments}
We have just scratched the tip of the iceberg. It will be important and interesting to understand the emergence of the consistency conditions that Polyakov has derived in the second half of his paper and see if they can be solved or understood even if some approximate sense--for example, can we understand the systematics of higher order or higher spin operators? This approach seems to have its own merit as we have demonstrated through the $\epsilon$-expansion. Apart from this, it will be interesting to reinterpret Polyakov's work \cite{polyakov} in terms of the modern bootstrap language. It will be also interesting to extend the formalism to other dimensions \cite{klebanov}. The success at order $O(\e^2)$ motivates the setting up bootstrap in the Mellin space \cite{rg, mack} which in some sense is the analog of the momentum space for conformal field theories. The non-perturbative nature of the formalism can be used for classification of CFTs in dimensions greater than two.

\section*{Acknowledgments} We thank B. Ananthanarayan, Rohini Godbole, Apratim Kaviraj, Zohar Komargodski,  Apoorva Patel and especially Joao Penedones  and Slava Rychkov for discussions. KS thanks Weizmann Institute of Science, Israel, Swansea University, Oxford University, Porto University, CERN, Geneva University, TU Vienna and ETH, Zurich for hospitality during the course of this work. AS acknowledges support from a Swarnajayanti fellowship, Govt. of India.

\appendix

\section{Determination of $T_d(p,q)$ in $4-\e$ dimensions}\label{A0}

We will now proceed with the determination of $T_d(p,q)$ in $4-\e$ dimensions as follows: This will be largely based on the analysis in appendix A of \cite{polyakov} except that there the result was presented in 4 dimensions. We will also stick to the sign conventions as in appendix A of \cite{polyakov} but in the main text the sign of $q$ will be reversed to be compatible with the direction of momenta in figure \eqref{fig:scatteringplot}. The momentum space representation of the three point vertex can be written as,
\be
\text{F.T.}\la T\mo(R)\phi(r)\phi(0)\ra=\int d^{4-\e}r d^{4-\e}R\frac{ e^{ipr+iqR}}{r^{2\D-d}R^d|R-r|^d}\,,
\ee
where $\D$ is the conformal dimension of $\phi$ and $d$ is the conformal dimension of the operator $\mo$. Using the Schwinger parameterization we can write,
\be\label{sch}
\frac{1}{a^s}=\frac{1}{\G(s)}\int_0^\infty dx\ x^{s-1}e^{-a x}\,.
\ee
Thus the above three point function in the momentum space becomes,
\be
\text{F.T.}\la T\mo(R)\phi(r)\phi(0)\ra=\text{const.}\ \int dx\ dy\ dz (xy)^{d/2-1}z^{\D-d/2-1} \int d^{4-\e}rd^{4-\e}R\ e^{f(x,y,z,r,R)}\,,
\ee 
where the function $f$ is given by,
\be
f(x,y,z,r,R)=ipr+iqR-x R^2-y (R-r)^2-z r^2\,.
\ee
What we do in the beginning is complete the squares for the function $f$ and isolate the parts depending solely on $r$ and $R$. After completing the squares, we get,
\be
f(x,y,z,r,R)=-A^2-B^2-f_1(x,y,z)\,,
\ee
where again the functions are given by,
\begin{align}\label{abf}
\begin{split}
A&=\sqrt{y+z}r+\frac{2yR-ip}{2\sqrt{y+z}}\,,\\
B&=\sqrt{\frac{xy+yz+xz}{y+z}}R+i\frac{(p+q)y+q z}{\sqrt{(y+z)(xy+yz+xz)}}\,,\\
f(x,y,z)&=\frac{p^2 x+q^2 z+(p+q)^2 y}{4(xy+yz+xz)}\\
\end{split}
\end{align}
We can first give a shift to the functions $A$ and $B$ and then carry out the integrals over the shifted variables by noticing that the integral measure is given by,
\be
\int d^{4-\e}r\rightarrow 4\pi \int r^{3-\e}dr\,.
\ee
Thus the integrals over $r$ and $R$ contribute to,
\be
\int d^{4-\e}rd^{4-\e}R e^{-(A^2+B^2)}=4\pi^2\G(2-\e/2)^2 (xy+yz+xz)^{-2+\e/2}\,,
\ee
and thus,
\be\label{ft}
\text{F.T.}\la T\mo(R)\phi(r)\phi(0)\ra=\int dxdydz\ \frac{(xy)^{d/2-1}z^{\D-d/2-1}}{(xy+yz+xz)^{2-\e/2}}e^{-f(x,y,z)}\,.
\ee
Replacing $(x,y,z)\rightarrow 1/4(1/x,1/y,1/z)$ gives,
\be
\text{F.T.}\la T\mo(R)\phi(r)\phi(0)\ra=\int dxdydz\ \frac{(xy)^{1-d/2-\e/2}z^{d/2+1-\D-\e/2}}{(x+y+z)^{2-\e/2}}e^{-\frac{xy q^2+yz p^2+xz(p+q)^2}{(x+y+z)}}\,.
\ee
We will now change to the coordinates $z=\n\r$, $x=\r\l$ and $y=\r(1-\l)$ and the integration limits are $0<\n,\r<\infty$ and $0<\l<1$ respectively. Thus the function in the argument of the exponential changes from,
\be\label{fgh}
f(\r,\n,\l)=-\frac{\r}{1+\n}(g(\l)+\n(h(\l))\,, \ \ g(\l)=q^2\l(1-\l)\,, \ \ h(\l)=p^2\l+(1-\l)(p+q)^2\,.
\ee
In these coordinates the integral takes the form,
\be
\int d\r d\n d\l\ \r^2 \frac{(\r^2\l(1-\l))^{1-\e/2-d/2}(\n\r)^{d/2+3-\D-\e/2}}{(\r(1+\n))^{2-\e/2}}e^{-f(\n,\l)\r}\,,
\ee 
where $f(\n,\l)$ is the remaining part of the argument in the exponential apart from $\r$. Performing the integral over $\r$ we get,
\be
\int_0^1 d\l\ (\l(1-\l))^{1-d/2-\e/2}\int_0^\infty d\n\ \frac{\n^{d/2+1-\D-\e/2}(1+\n)^{2-\e/2-d/2-\D}}{[g(\l)+\n h(\l)]^{4-\e-d/2-\D}}\,.
\ee
Finally performing the integral over $\n$ we get,
\begin{align}
\text{F.T.}\la T\mo(R)\phi(r)\phi(0)\ra&=N_d T_{4-\e-d}(p,q)+L_d T_{d}(p,q)q^{2d-4+\e}\,,
\end{align}
where,
\be
N_d=\frac{\G(\D-d/2)\G(d-2+\e/2)}{\G(\D+d/2-2+\e/2)}\,, \ \ L_d=\frac{\G(2+d/2-\D-\e/2)\G(2-d-\e/2)}{\G(4-d/2-\D-\e)}
\ee
\begin{align}
T_d(p,q)&=\int_0^1d\l\  \frac{(\l(1-\l))^{d/2-1}}{[\l p^2+(1-\l)(p+q)^2]^{2+d/2-\D-\e/2}}\nonumber\\
&\times {}_2F_1\bigg[2+d/2-\D-\e/2,\D+d/2-2+\e/2,d-1+\e/2,\frac{q^2\l(1-\l)}{\l p^2+(1-\l)(p+q)^2}\bigg]\,.
\end{align}
In the expression for the three point functions we will take the discontinuous part which amounts to taking the part proportional to $q^{2d-4+\e}$. 

\section{ Mixing coefficients}\label{mixapp}
We will derive the mixing factors for the channels below. The full scattering amplitude is given by
\be
A^{(s)}_{ijkl}=a_0 T^0(q,p)\text{Im}D^0(q) T^0(q,p')\d_{ij}\d_{kl}+a_2 T^2(q,p)\text{Im}D^2(q) T^2(q,p')(\d_{ik}\d_{jl}+\d_{il}\d_{jk}-\frac{2}{n}\d_{ij}\d_{kl})\,.
\ee
where the labels $I=0,2$ denote the contribution of the exchanged scalars $O^{(0)}$ and $O^{(2)}$ in the theory. The $t$-channel is obtained by switching $j\leftrightarrow k$ and the $u$-channel is obtained by switching $j\leftrightarrow l$ respectively. Thus
\begin{align}
\begin{split}
A^{(t)}_{ijkl}&=a_0 T^0(q,p)\text{Im}D^0(q) T^0(q,p')\d_{ik}\d_{jl}+a_2 T^2(q,p)\text{Im}D^2(q) T^2(q,p')(\d_{ij}\d_{kl}+\d_{il}\d_{jk}-\frac{2}{n}\d_{ik}\d_{jl})\,,\\
A^{(u)}_{ijkl}&=a_0 T^0(q,p)\text{Im}D^0(q) T^0(q,p')\d_{il}\d_{jk}+a_2 T^2(q,p)\text{Im}D^2(q) T^2(q,p')(\d_{ik}\d_{jl}+\d_{ij}\d_{kl}-\frac{2}{n}\d_{il}\d_{jk})\,.
\end{split}
\end{align}
Thus multiplying the total amplitude $A_{ijkl}=A^{(s)}_{ijkl}+A^{(t)}_{ijkl}+A^{(u)}_{ijkl}$ by $\d^{ij}\d^{kl}$ we have,
\begin{align}
\begin{split}
A_{ijkl}\d^{ij}\d^{kl}&=n^2\bigg[a_0\bigg(1+\frac{2}{n}\bigg)T^0 \text{Im}D^0(q)T^0+\frac{2n^2+2n-4}{n^2}a_2 T^2 \text{Im}D^2(q)T^2\bigg]\,,\\
A_{ijkl}\bigg(\d^{ik}\d^{jl}+\d^{il}\d^{jk}-\frac{2}{n}\d^{ij}\d^{kl}\bigg)&=(2n^2+2n-4)\bigg[a_2T^2 \text{Im}D^2(q)T^2+a_0T^0 \text{Im}D^0(q)T^0\\
&+\frac{n-2}{n}a_2T^2 \text{Im}D^2(q)T^2\bigg]\,.
\end{split}
\end{align}
In the limit where the $t$ and the $u$ channel contributes, the $\d^i$ dependent factors drop out from both $T^0 \text{Im}D^0(q)T^0$ and $T^2 \text{Im}D^2(q)T^2$. Thus we can write the above expression as,
\be
A_{ijkl}=\bigg[a_I+\sum_{J=0,2}c_{IJ}a_J\bigg]f(v)\,.
\ee
Comparing with the above formula for $A_{ijkl}$ we get,
\begin{align}\label{mixcoeff}
\begin{split}
c_{00}&=\frac{2}{n}\,,\ \ c_{02}=\frac{2n^2+2n-4}{n^2}\,,\\
c_{20}&=1\,,\ \ c_{22}=\frac{n-2}{n}\,.
\end{split}
\end{align}

\section{ Evaluation of some generic integrals}\label{A}
In the evaluation above we will be facing the integrals of two generic types. We digress a little to show the evaluation of these integrals. First note that the four point amplitudes in the momentum space involve terms of the form,
\be
A(q,p,p')\sim q^a p^b p'^c\,.
\ee
The fourier transform of these power law behaviour in the regime of interest is not very difficult to evaluate. 
\be
\int_0^{R^{-1}} d^{4-\e}q\int_{R^{-1}}^{r'^{-1}} \frac{dp'}{p'}p'^{-\e}\bigg(\int_{p'}^{1/r}+\int_{p'}^{1/r'}\bigg)\frac{dp}{p}p^{-\e}q^a p^b p'^c\,.
\ee
Using the Euclidean representation for the integral over $q$ we can show that,
\be
\int_0^{R^{-1}}d^{4-\e}q\ q^a=\frac{1}{4+a-\e}R^{\e-4-a}\,.
\ee
For the remaining part, the integral over $p$ and $p'$ gives,
\begin{align}
\begin{split}
\frac{1}{b-\e}\int_{R^{-1}}^{r'^{-1}}\frac{dp'}{p'}p'^{c-\e}(r^{\e-b}+r'^{\e-b}-2p'^{b-\e})=&
\frac{1}{b-\e}\bigg[\frac{1}{c-\e}(r^{\e-b}+r'^{\e-b})(r'^{\e-c}-R^{\e-c})\\
&-\frac{2}{b+c-2\e}(r'^{2\e-b-c}-R^{2\e-b-c})\bigg]\,.
\end{split}
\end{align}

Combining this together we have for $b\neq\e$,
\begin{align}
\begin{split}
&S_{a,b,c}(r,r',R)=\int_0^{R^{-1}} d^{4-\e}q\int_{R^{-1}}^{r'^{-1}} \frac{dp'}{p'}p'^{-\e}\bigg(\int_{p'}^{1/r}+\int_{p'}^{1/r'}\bigg)\frac{dp}{p}p^{-\e}q^a p^b p'^c\\
&=\frac{R^{\e-4-a}}{(b-\e)(4+a-\e)}\bigg[\frac{1}{c-\e}(r^{\e-b}+r'^{\e-b})(r'^{\e-c}-R^{\e-c})\\
&-\frac{2}{b+c-2\e}(r'^{2\e-b-c}-R^{2\e-b-c})\bigg]\,.
\end{split}
\end{align}
When $b=\e$, the integral over $p$ and $p'$ gives,
\begin{align}
\begin{split}
&T_{a,c}(r,r',R)=\int_0^{R^{-1}} d^{4-\e}q\int_{R^{-1}}^{r'^{-1}} \frac{dp'}{p'}p'^{-\e}\bigg(\int_{p'}^{1/r}+\int_{p'}^{1/r'}\bigg)\frac{dp}{p}p^{-\e}q^a p^\e p'^c\\
&=\frac{R^{\e-4-a}}{(4+a-\e)(c-\e)}\bigg[\log \bigg(\frac{1}{r r'}\bigg)(r'^{\e-c}-R^{\e-c})-2(R^{\e-c}\log R-r'^{\e-c}\log r')\\
&+\frac{2}{c-\e}(r'^{\e-c}-R^{\e-c})\bigg] \,.
\end{split}
\end{align}

\section{ Generic three point function}\label{B}

To calculate the three point functions of the form $\la \phi_i(r)\phi_j(0)(J_\m)_{kl}(R)\ra$ we will need to evaluate some generic integrals in the momentum space of the form,
\be
\text{F.T.}\la T\mo_{kl}(R)\phi_i(r)\phi_j(0)\ra={\mathcal{I}}_{ijkl}\int d^{4-\e}r d^{4-\e}R \frac{e^{i(pr+qR)}}{|R-r|^a R^b r^c}\,,
\ee
where ${\mathcal{I}}_{ijkl}$ is the associated $O(N)$ tensor structure (warning: here $a$ is not $4-\e$!). Using the Schwinger parameterization of the propagators as in \eqref{sch}, we can get an integral similar to \eqref{ft} (modulo the tensor structure)
\be
 \text{F.T.}\la T\mo_{kl}(R)\phi_i(r)\phi_j(0)\ra=\text{const}\cdot \int dxdydz\ \frac{x^{a/2-1}y^{b/2-1}z^{c/2-1}}{(xy+yz+xz)^{2-\e/2}}e^{-f(x,y,z)}\,.
\ee
where $f(x,y,z)$ has the exact same structure as in \eqref{abf}. We can now do the analogous change of variables from $(x,y,z)\rightarrow 1/4(1/x,1/y,1/z)$ to get,
\be
 \text{F.T.}\la T\mo_{kl}(R)\phi_i(r)\phi_j(0)\ra=\text{const}\cdot \int dxdydz\ \frac{x^{1-a/2-\e/2}y^{1-b/2-\e/2}z^{1-c/2-\e/2}}{(x+y+z)^{2-\e/2}}e^{-f(1/x,1/y,1/z)}\,.
\ee
Changing the variables to,
\be
x=\r\l\,, \ \ y=\r(1-\l)\,,\ \ \text{and} \ \ z=\r\n\,,
\ee
we can put the integral above in the form,
\begin{align}
\begin{split}
 \text{F.T.}\la T\mo_{kl}(R)\phi_i(r)\phi_j(0)\ra&=\text{const}. \int_0^\infty \r^{3-\frac{a+b+c}{2}-\e}e^{-f(\n\l)\r}\times\int d\l d\n \l^{1-\frac{a}{2}-\frac{\e}{2}}(1-\l)^{1-\frac{b}{2}-\frac{\e}{2}}\frac{\n^{1-\frac{c}{2}-\frac{\e}{2}}}{(1+\n)^{2-\frac{\e}{2}}}\,,\\
 &=\text{const}.\int_0^1 d\l\ \l^{1-\frac{a}{2}-\frac{\e}{2}}(1-\l)^{1-\frac{b}{2}-\frac{\e}{2}}\int_0^\infty d\n\ \frac{\n^{1-\frac{c}{2}-\frac{\e}{2}}(1+\n)^{2-\frac{\e}{2}-\frac{a+b+c}{2}}}{[g(\l)+\n h(\l)]^{4-\e-\frac{a+b+c}{2}}}\,,\\
&=\text{const}.[N_{a,b,c}T^1_{a,b,c}(p,q)+L_{a,b,c}T^2_{a,b,c}(p,q)s^{-2+\frac{a+b+\e}{2}}]\,.
\end{split}
\end{align}
where $g(\l)$ and $h(\l)$ are defined in \eqref{fgh} and the two constants take the form,
\be\label{NL}
N_{a,b,c}=\frac{\G(\frac{c}{2})\G(-2+\frac{\e}{2}+\frac{a+b}{2})}{\G(-2+\frac{a+b+c+\e}{2})}\,,\ \ L_{a,b,c}=\frac{\G(2-\frac{\e}{2}-\frac{a+b}{2})\G(2-\frac{c}{2}-\frac{\e}{2})}{\G(4-\e-\frac{a+b+c}{2})}\,.
\ee
The functions $T^1$ and $T^2$ are given by,
\begin{align}\label{t1t2}
\begin{split}
T^1_{a,b,c}&=\int d\l\ \frac{\l^{1-\frac{a}{2}-\frac{\e}{2}}(1-\l)^{1-\frac{b}{2}-\frac{\e}{2}}}{h(\l)^{4-\frac{a+b+c}{2}-\e}}{}_2F_1\bigg[\frac{c}{2},4-\e-\frac{a+b+c}{2},3-\frac{\e}{2}-\frac{a+b}{2},\frac{g(\l)}{h(\l)}\bigg]\,,\\
T^2_{a,b,c}&=\int d\l\ \frac{\l^{\frac{b}{2}-1}(1-\l)^{\frac{a}{2}-1}}{h(\l)^{2-\frac{c+\e}{2}}}{}_2F_1\bigg[2-\frac{c+\e}{2},-2+\frac{a+b+c+\e}{2},-1+\frac{a+b+\e}{2},\frac{g(\l)}{h(\l)}\bigg]\,,\\
\end{split}
\end{align}
where $g(\l)$ and $h(\l)$ are given in \eqref{fgh}.

\section{ Four point Green function of the mixed correlator}\label{C}

We quote the relevant details for the construction of the four point Green function for the mixed correlator in the position space. 

\subsection{Unitary Green function}\label{5}
 The way to schematically write down the four point function in momentum space, is,
\be
F.T\{\la T \phi_i(r)\phi_j(0)\phi_k(R)\phi^2\phi_l(R+r')\ra\}=A_I(q,p,p')=\int_s^\infty \frac{ds'}{s'}T^2_{\phi\phi\mo}(s',p)\text{Im}D(s') T^2_{\phi(\phi^2\phi)\mo}(s',p')\,,
\ee
where the associated three point functions relevant for the leading discontinuous part of the amplitude is given by $T^2_{a,b,c}$ functions of the appendix \eqref{B}. We have also suppressed the indices associated with the four point amplitude for convenience. Later we will use superscripts $s,t,u$ to denote the contributions due to the individual channels. Since we have already defined the generic three point functions for scalars, we can refer to \eqref{t1t2} of \eqref{B} for this section. We will need two different kinds of three point functions for our purposes. These are defined as $F_1=T^2_{d,d,2\D-d}$ and $G_1=T^2_{d+\D_3-\D,d+\D-\D_3,\D+\D_3-d}$. Thus the four point amplitude in the momentum space becomes,
\be
A_I(q,p,p')=\int_s^\infty \frac{ds'}{s'}F_1(s',p)\ \text{Im}D(s')\ G_1(s',p')\,.
\ee
This four point function should now be solved in the regime $p\gg p'\gg q$ for the cases when $s\gg (v,w)$ and $s\ll (v,w)$ respectively. We will now proceed to evaluate the limits for the individual functions $F_1$ and $G_1$ separately. For the function $F_1$, we know the limits from our previous calculations. Barring the overall factors, the expression for $F_1$ in the two limits take the form,
\be
F_1(s,v)=\frac{1}{v^2} \left\{
     \begin{array}{lr}
      f_1 v^{-d/2+\D+\e/2} & : s\ll v\\
      f_2 s^{2-d/2-\D-\e/2}v^{-2+2\D+\e} & : s\gg v
     \end{array}
   \right.
\ee
where the coefficients $f_1$ and $f_2$ can be written as,
\be
f_1=\frac{\G(d/2)^2}{\G(d)}\,, \ \ f_2=\frac{\G(2-\D-\frac{\e}{2})^2\G(d-1+\frac{\e}{2})}{\G(1+\frac{d}{2}-\D)\G(2+\frac{d}{2}-\D-\frac{\e}{2})}\text{Re}[(-1)^{\frac{1}{2}(-d-2\D-\e)}]\,.
\ee

And for the other function we have,
\be
G_1(s,v)=\frac{1}{v^2}\left\{
\begin{array}{lr}
g_1 v^{\frac{\D+\D_3-d+\e}{2}}& : s\ll v\\
g_2^{(0)} s^{-2-\frac{d}{2}+\frac{\D+\D_3}{2}+\frac{\e}{2}} v^2+g_2^{(1)} s^{2-\frac{d}{2}-\frac{\D+\D_3}{2}-\frac{\e}{2}}v^{-2+\D+\D_3+\e}& :s\gg v
\end{array}
\right.
\ee
where the coefficients are,
\begin{align}
\begin{split}
g_1&=\frac{\G((d+\D-\D_3)/2)\G((d+\D_3-\D)/2)}{\G(d)}\,,\\
g_2^{(0)}&=\frac{\G(d-1+\frac{\e}{2})\G(-2+\D+\frac{\e}{2})\G(-2+\D_3+\frac{\e}{2})}{\G(-2+\frac{d+\D+\D_3+\e}{2})\G(-3+\frac{d+\D+\D_3}{2}+\e)}\text{Re}[(-1)^{\frac{1}{2}(-d+\D+\D_3+\e)}]\,,\\
g_2^{(1)}&=\frac{\G(2-\D-\frac{\e}{2})\G(2-\D_3-\frac{\e}{2})\G(d-1+\frac{\e}{2})}{\G(1+\frac{d-\D-\D_3}{2})\G(2+\frac{d-\D-\D_3-\e}{2})}\text{Re}[(-1)^{\frac{1}{2}(-d-\D-\D_3-\e)}]\,.
\end{split}
\end{align}
In addition to the above information, we also have for the exchange,
\be
\text{Im} D(s)=\text{const}.\ s^{d-a/2}\,.
\ee
We can thus break up the integral over $s'$ into regimes 
\be
A(q,p,p')=\int_s^w \frac{ds'}{s'}(\cdots)+\int_w^v \frac{ds'}{s'}(\cdots)+\int_v^\infty \frac{ds'}{s'}(\cdots)\,,
\ee
where the terms in $(\cdots)$ denote the function $F_1\times \text{Im}D(s)\times G_1$ in various limits and $v=p^2$, $s=q^2$ and $w=p'^2$ respectively. We can spell out the different individual limits for the convenience of the reader. 

\begin{itemize}
\item $s\ll v,w$. In this case, we find that,
\be
F_1(s,v)\ \text{Im}D(s)\ G_1(s,w)=\frac{1}{v^2w^2} f_1 g_1 v^{-d/2+\D+\e/2}w^{\frac{\D+\D_3-d+\e}{2}}s^{d-a/2}\,.
\ee

\item $w\ll s\ll v$. In this case, we find that,
\begin{align}
\begin{split}
F_1(s,v)\ \text{Im}D(s)\ G_1(s,w)=&\frac{1}{v^2w^2} f_1 v^{-d/2+\D+\e/2}s^{d-a/2}[g_2^0 s^{-2-\frac{d}{2}+\frac{\D+\D_3}{2}+\frac{\e}{2}} w^2\\
&+g_2^1 s^{2-\frac{d}{2}-\frac{\D+\D_3}{2}-\frac{\e}{2}}w^{-2+\D+\D_3+\e} ] \,.
\end{split}
\end{align}

\item $s\gg v,w$. In this case, we find,
\begin{align}\label{sggv}
\begin{split}
F_1(s,v)\ \text{Im}D(s)\ G_1(s,w)=&\frac{1}{v^2w^2} f_2 s^{2+d/2-a/2-\D-\e/2}v^{-2+2\D+\e}[g_2^0 s^{-2-\frac{d}{2}+\frac{\D+\D_3}{2}+\frac{\e}{2}} w^2\\
&+g_2^1 s^{2-\frac{d}{2}-\frac{\D+\D_3}{2}-\frac{\e}{2}}w^{-2+\D+\D_3+\e} ] \,.
\end{split}
\end{align}

\end{itemize}

We are now supposed to complete the integral over $s'$ to get the four point amplitude $A'_{ijkl}(q,p,p')$. Modulo the overall coefficients which are present and the factors due to the $s'$ integral, we can write this as a function of four different terms 
\begin{align}\label{mixeds}
\begin{split}
A^{(s)}_I(q,p,p')=&\frac{a_I}{v^2w^2}[c_1 s^{d-a/2}v^{-d/2+\D+\e/2}w^{(\D+\D_3-d+\e)/2}+c_2 v^{-d/2+\D+\e/2}w^{(\D+\D_3+d-a+\e)/2}\\
&+c_3 v^{-2-a/2+(3\D+\D_3)/2+\e}w^2+c_4v^{2-a/2+(\D-\D_3)/2}w^{-2+\D+\D_3+\e}]\,,
\end{split}
\end{align}
where the coefficients in the above expression are given by,
\begin{align}\label{ci}
\begin{split}
c_1&=-\frac{f_1g_1}{d-a/2}\,,\\ 
c_2&=\frac{f_1g_1}{d-a/2}-2\bigg[\frac{f_1g_2^{(0)}}{d-4-a+\e+\D+\D_3}+\frac{f_1g_2^{(1)}}{d+4-a-\e-\D-\D_3}\bigg]\,,\\
c_3&=\frac{2f_1g_2^{(0)}}{d-4-a+\e+\D+\D_3}-\frac{2f_2g_2^{(0)}}{\D_3-\D-a}\,,\\ c_4&=\frac{2f_1g_2^{(1)}}{d+4-a-\e-\D-\D_3}-\frac{f_2g_2^{(1)}}{4-a/2-\e-(\D_3+3\D)/2}\,.
\end{split}
\end{align}
We have omitted the overall tensor structure on the \textit{rhs} which is to say that these functions will depend on the kind of exchange we have and that will reflect through the individual coefficients $c_1^I$ etc. appearing above. The superscript $(s)$ denotes that the above expressions are for the $s-$ channel decomposition of the amplitude. The $t$ and the $u$ channel will only have the last limit \textit{viz.} $s\gg v,w$ and hence only the expression \eqref{sggv} in the decomposition albeit with some different coefficients. Thus for the $t$ and the $u$ channel there is a change of the variables for the three point functions. For the $\la\phi\phi\mo\ra$ correlator notice that, the function $T^2_d(p,q)\rightarrow T^2_d(p-q,p-p')$ and similarly for the other three point function $\la\phi(\phi^2\phi) \mo\ra$. Notice also that the exchange momentum is now $s=(p-p')^2$ and for the regime $P\gg p'\gg q$, $s\geq v$. Thus for the $t$ channel only one regime is relevant \textit{viz.} $s\gg v,w$. So it will suffice to calculate the three point functions in this regime. The respective three point functions are given by, 
\begin{align}
\begin{split}
\la \phi\phi\mo\ra&=\frac{1}{v^2w^2}h_1 s^{2-d/2-\D-\e/2} (v w)^{\D+\e/2}\ :s\gg v,w\,,\\
\la \phi(\phi^2\phi)\mo\ra&=h_2 s^{2-\frac{d}{2}-\frac{\D_3+\D}{2}-\frac{\e}{2}} v^{\D_+\frac{\e}{2}-2} w^{-2+\D_3+\frac{\e}{2}}+h_3 s^{-2-\frac{d}{2}+\frac{\D+\D_3}{2}+\frac{\e}{2}}\ :s\gg v,w\,.
\end{split}
\end{align}
where the explicit form of the coefficients $h_1$, $h_2$ and $h_3$ are given by,
\begin{align}
\begin{split}
h_1&=\frac{\G(2-\D-\frac{\e}{2})^2\G(d-1+\frac{\e}{2})}{\G(1+\frac{d}{2}-\D)\G(2+\frac{d}{2}-\D-\frac{\e}{2})}\text{Re}[(-1)^{-\frac{1}{2}(d+2\D+\e)}]\,,\\
h_2&=\frac{\G(2-\D-\frac{\e}{2})\G(2-\D_3-\frac{\e}{2})\G(d-1+\frac{\e}{2})}{\G(1+\frac{d-\D-\D_3}{2})\G(2+\frac{d-\D-\D_3-\e}{2})}\text{Re}[(-1)^{-\frac{1}{2}(d+\D+\D_3+\e)}]\,,\\
h_3&=\frac{\G(d-1+\frac{\e}{2})\G(-2+\D+\frac{\e}{2})\G(-2+\D_3+\frac{\e}{2})}{\G(-2+\frac{d+\D+\D_3+\e}{2})\G(-3+\e+\frac{d+\D+\D_3}{2})}\text{Re}[(-1)^{\frac{1}{2}(-d+\D+\D_3+\e)}]\,.
\end{split}
\end{align}
Similarly, for the $u$ channel, the exchange momentum is $s=(p+p'-q)^2$ which for our regime of interest is again $s\geq v$. Hence the above limits of the three point functions apply in this channel as well. The full four point function can be written as,
\be
A^{(t)}_I(s,v,w)=\la \phi_i \phi_j \mo\ra\ \text{Im}D(s)\ \la \phi_k (\phi^2\phi_l) \mo\ra\,.
\ee
 Integrating over the variable $s$ in the regime $s\gg v\gg w$ one can see that,
\begin{align}\label{momt}
\begin{split}
A^{(t)}_I(v,w)=\int_v^\infty\frac{ds'}{s'}A^{(t)}_I(s',v,w)=&\frac{2}{v^2w^2}\bigg[\frac{h_1h_2}{-8+a+3\D+\D_3+2\e}v^{2-\frac{a}{2}+\frac{\D-\D_3}{2}}w^{-2+\D+\D_3+\e}\\
&+\frac{h_1h_3}{a+\D-\D_3}v^{-\frac{a}{2}+\frac{\e}{2}+\frac{\D+\D_3}{2}}w^{\D+\frac{\e}{2}}\bigg]\,.
\end{split}
\end{align}
We are now in a position to address the question about whether this mixed correlator can determine the anomalous dimension for the operators $\phi_i$ and the composite operator $\phi^2\phi_i$ given the information about the anomalous dimensions for the exchange operators. The key idea will be now to convert the sum of the contributions from the three channels into position space representation and impose the operator algebra which we now proceed to do in the next section. 

\vspace{5mm}

\subsection{ Position space representation}

The momentum space amplitude can now be converted into the position space through the Fourier transform so that we can impose the operator algebra (OPE) on that. We will consider individual terms in individual channels for clarity.

\subsubsection{$s$ channel}
 Consider the $s$ channel first in \eqref{mixeds}. In terms of $v=p^2$, $w=p'^2$ and $s=q^2$ this becomes,
\begin{align}\label{moms}
\begin{split}
A^s_{I}(q,p,p')=&\frac{a_I}{p^4p'^4}[c_1^I q^{2d-a}p^{-d+2\D+\e}p'^{\D+\D_3-d+\e}+c_2^Ip^{-d+2\D+\e}p'^{\D+\D_3+d-a+\e}\\
&+c_3^I p^{-4-a+\D_3+3\D+2\e}p'^4+c_4^I p^{4-a-\D_3+\D}p'^{-4+2(\D+\D_3)+2\e}]\,.
\end{split}
\end{align}
The label $i$ indicates what scalar ($i=0,2$) is exchanged in this channel. The fourier transform of the above function is,
\be
G_{I}(r,r',R)=\int d^{4-\e}p\ d^{4-\e}p' d^{4-\e}q\ e^{i (pr+p'r'+q R)}A_{I}(q,p,p')\,.
\ee
In the regime we are interested, we can simplify the measure further by writing,
\begin{align}
\begin{split}
\int\frac{1}{p^4p'^4} d^{4-\e}p\ d^{4-\e}p' d^{4-\e}q\ e^{i (pr+p'r'+q R)}(\cdots)&\equiv\int_0^{R^{-1}} q^{3-\e}dq\ \int_{R^{-1}}^{r'^{-1}} \frac{dp'}{p'}p'^{-\e}\int_{p'}^{r^{-1},r'^{-1}}\frac{dp}{p}p^{-\e}(\cdots)\\
&\equiv \text{FT}(\cdots)\,.
\end{split}
\end{align}
We refer the reader to the appendix \ref{A} for the details of the Fourier transforms and quote here the final results. We are interested in the leading power law dependence of the position space Green functions after the matching with the OPE part. Now we are in a position to match various coefficients. First notice that in the position space representation, two terms appear repeatedly. These are,
\be
r'^{a-3\D-\D_3}\ \ \text{and}\ \ R^{a-3\D-\D_3}\,.
\ee
But for the regime of interest, we have chosen $r\sim r'\ll R$ and even at the leading order we know that $a-3\D-\D_3=-2+O(\e)$. Thus the dominant contribution in the specific regime will come from $r^{a-3\D-\D_3}$ term in the position space representations. We can thus compare the coefficients of this particular term in the position space representation of the Green function. The coefficients for this term is,
\begin{align}
\begin{split}
-\frac{a_I}{a(a-3\D-\D_3)}\bigg[\frac{(\D-\D_3-2d_I+a)}{(\D+\D_3+d_I-a)(2\D-d_I)}c_2^I&+\frac{3\D+\D_3-3a}{a(3\D+\D_3-2a)}c_3^I
\\
&+\frac{a-\D-3\D_3}{(\D-\D_3)(2(\D+\D_3)-a)}c_4^I\bigg]\,.
\end{split}
\end{align}

\subsubsection{$t/u$ channel}
Fourier transforming the $t/u$ channel contributions into position space, we find that the leading contribution to the $r'^{a-3\D-\D_3}$ comes with the coefficient,

\begin{align}\label{alphai}
\begin{split}
\a&=\frac{1}{a(a-3\D-\D_3)}\bigg[\frac{h_1 h_3}{\D(-a+\D+\D_3)}\\
&-\frac{2h_1h_2(-8+a+\D+3\D_3+2\e)}{(-4+a-\D+\D_3+\e)(2(-2+\D+\D_3)+\e)(-8+a+3\D+\D_3+2\e)}\bigg]
\end{split}
\end{align}

Taking the coefficients of $r'^{a-3\D-\D_3}$ from the position space representation of the Green function we get for each exchange ($I=0,2$),
\begin{align}\label{constraint}
\begin{split}
0=\sum_{J,=0,2}c_{IJ}\a_Ja_J-\frac{a_I}{a(a-3\D-\D_3)}\bigg[&\frac{(\D-\D_3-2d_I+a)}{(\D+\D_3+d_I-a)(2\D-d_I)}c_2^I+\frac{3\D+\D_3-3a}{a(3\D+\D_3-2a)}c_3^I\\
&+\frac{a-\D-3\D_3}{(\D-\D_3)(2(\D+\D_3)-a)}c_4^I\bigg]\,.
\end{split}
\end{align}
The coefficients $\a_I$ will contain information about the mixing in the $t$ and $u$ channel as given in \eqref{alphai}, while $c_{IJ}$ are the mixing coefficients given in \eqref{mixcoeff} and the other coefficients are given in \eqref{ci}. The coefficients $a_i$ contains the product of the OPE coefficients coming from two different OPE one with $\phi_i\times\phi_j$ and other with $\phi_i\times (\phi^2\phi_j)$.  
With this it should be possible to calculate the anomalous dimensions for the operators $\phi_i$ and $\phi^2\phi_i$ provided we know the anomalous dimensions of the exchanges upto required order in $\e$.

\end{document}